\documentclass[aps,prl,reprint,twocolumn,groupedaddress,longbibliography,nobibnotes,nofootinbib,floatfix]{revtex4-2}
\usepackage{amsmath,amssymb,amsthm,bm,mathtools,empheq}
\usepackage{microtype}
\usepackage{float}
\usepackage{xcolor}
\usepackage{tikz}
\usetikzlibrary{arrows.meta,positioning}
\colorlet{MAGENTA}{magenta}
\usepackage[colorlinks=true,citecolor=blue,linkcolor=magenta,urlcolor=teal]{hyperref}

\newcommand{\Tr}{\operatorname{Tr}}

\newcommand{\EE}{\mathbb{E}}
\newcommand{\RR}{\mathbb{R}}
\newcommand{\Id}{\mathbb{I}}
\newcommand{\T}{^{\mathsf T}}
\newcommand{\bx}{{\boldsymbol x}}
\newcommand{\by}{{\boldsymbol y}}
\newcommand{\bxi}{{\boldsymbol \xi}}

\newcommand{\Ker}{\mathcal{C}}

\newcommand{\PhiTot}{\Phi_{\textrm{total}}}

\DeclareRobustCommand{\tcOld}[1]{\textcolor{magenta}{#1}}
\DeclareRobustCommand{\tcNew}[1]{\textcolor{blue}{#1}}
\newcommand{\tcRedDispatch}[1]{\tcNew}
\newcommand{\tc}[1]{%
  \ifnum\pdfstrcmp{\detokenize{#1}}{2}=0
    \expandafter\tcRedDispatch
  \else
    \expandafter\tcOld
  \fi
  {#1}%
}

\newcommand{\mn}[1]{\textcolor{red}{#1}}

\newtheorem{theorem}{Theorem}

\begin{document}

\title{Directed walks shape a universal square-root law of entropy production rate in nonreciprocal systems}
%\title{Directed walks determine entropy production in nonreciprocal systems}
%\title{Square-Root Law of Entropy Production in Directed Networks}
%\title{\tc{Walk-resolved entropy production in nonreciprocal networks}}

\author{Thiparat Chotibut}
\email{thiparatc@gmail.com}
\affiliation{
Chula Intelligent and Complex Systems Center of Excellence, Department of Physics, Faculty of Science, Chulalongkorn University, Bangkok 10330, Thailand \\School of Physical and Mathematical Sciences, Nanyang Technological University, Singapore}
\author{Ewa Gudowska-Nowak}
\affiliation{Institute of Theoretical Physics and Mark Kac Center for Complex Systems Research, Jagiellonian University, 30-348 Krak\'ow, Poland}
\author{Maciej A. Nowak}
\affiliation{Institute of Theoretical Physics and Mark Kac Center for Complex Systems Research, Jagiellonian University, 30-348 Krak\'ow, Poland}
\date{\today}

% The Main Letter, End Matter, and bibliography are assembled in this file.
% =============================================================
% MAIN LETTER AND END MATTER -- inlined for one-file compilation
% =============================================================
\begin{abstract}
The entropy production rate (EPR) quantifies irreversibility of a nonequilibrium steady state, yet standard formulas obscure how a complex interaction network generates it. For multivariate Ornstein-Uhlenbeck dynamics on such networks, we express the EPR as a quadratic form in antisymmetric matrices measuring the nonreciprocity of aggregate directed walks at every length, and, equivalently, as two weighted-walk quantities: pairs of directed walks sharing both endpoints, and directed closed walks. For diagonalizable interactions, an exact correspondence translates these walk quantities into eigenvalues and biorthogonal eigenvector overlaps. Across dense, sparse, and deep acyclic random interactions satisfying matched-walk conditions, the mean EPR per node universally follows the \emph{square-root law} $\phi_*(g)=1-\sqrt{1-g^2}$, where $g \in [0,1)$ parametrizes the interaction strength. Deep acyclic interaction matrices are nilpotent, with all eigenvalues fixed at zero for every $g$, yet, as their depth increases, their mean EPR per node approaches $\phi_*(g)$. Thus, the square-root law arises from  directed walk properties, rather than from a shared spectral density or specific network topology.
\end{abstract}
\maketitle

\paragraph{Introduction.}---Fluctuations and dissipation around a stable state in a complex system are naturally described by multivariate Ornstein-Uhlenbeck (OU) dynamics~\cite{Risken1996Fokker,Gardiner2010Stochastic,Van1992Stochastic}. This stochastic process generically characterizes the steady states in systems with many interacting degrees of freedom, from neural circuits~\cite{Sompolinsky1988,Kadmon2015,Dayan2005,Rungratsameetaweemana2025}, coupled oscillators~\cite{Sakaguchi1988,Castellano2009}, and ecological communities~\cite{May1972,Bunin2017,Allesina2015,Biroli2018Marginally} to gene regulatory networks~\cite{Alon2006,Paulsson2005Models}. For isotropic noise, nonreciprocal interactions break detailed balance and the system fluctuates and dissipates in a nonequilibrium steady state (NESS), sustaining stationary probability currents~\cite{Zia2007,Qian2006,Schnakenberg1976Network,Lai2026noisy}.
%\footnote{Such nonreciprocal interactions also organize collective dynamics far beyond the OU setting~\cite{Fruchart2021,Loos2020Irreversibility} \tc{[add more references?]}.} 
The thermodynamic cost of sustaining these currents is the entropy production rate (EPR), a central quantity in stochastic thermodynamics~\cite{Seifert2012Stochastic,Falasco2024Macroscopic}. The EPR bounds current fluctuations through the thermodynamic uncertainty relation~\cite{Barato2015Thermodynamic,Gingrich,Horowitz2020Thermodynamic}, limits sensory precision~\cite{Lan2012Energy,Tostevin2009} and constrains the cost of learning~\cite{Goldt2017Stochastic}. It is also increasingly used as a noninvasive marker of biological function in living matter~\cite{Lynn2021broken,Gilson2023PRE,SanzPerl2021nonequilibrium,Battle2016,Gnesotto2018}.

Despite its central role, the EPR of OU dynamics is commonly obtained from the stationary covariance, typically by solving a large Lyapunov equation numerically, or from an equivalent spectral representation~\cite{Gardiner2010Stochastic,Tome2010,Godreche2019Characterising,Fyodorov2025Nonorthogonal}. These approaches are algorithmic and systematic, yet they do not explicitly reveal how the structure or the topology of the interaction network generates dissipation. Recent studies show that non-normality of the interaction network can correlate strongly with the EPR within particular  families~\cite{Kaluarachchi2024Broken,Nagayama2026Oscillatory}, yet non-normality is not always necessary: normal directed circulant matrices can generate a large EPR~\cite{Godreche2019Characterising,Kaluarachchi2024Broken}. These tensions, together with growing interest in stochastic dynamics in sparse networks~\cite{Metz2025Dynamical,MetzPerezCastillo2026DynamicalCavity,Neri2012,Metz2019Spectra,Chotibut2025,Rungratsameetaweemana2025}, call for an analytical framework linking the EPR to the structure of an arbitrary interaction network that sustains a NESS.

In this Letter, we show how the structure of an interaction network  $W \in \mathbb{R}^{N \times N}$ shapes its EPR under stable OU dynamics with a nondegenerate Gaussian white noise, presenting \emph{four main results}. First, we express the EPR as an exact quadratic form in the $r$-step walk-nonreciprocity matrices 
\begin{equation}\label{eq:A-def}
\mathcal A^{(r)}:=\tfrac12[W^r-(W^r)\T],
\end{equation} 
whose $(i,j)$ entry reflects the difference between the aggregate weights of all $r$-step walks from $j$ to $i$ and from $i$ to $j$; $\mathcal A^{(r)}$ vanishes identically when aggregate $r$-step propagation is reciprocal between every pair of nodes. Each pair of these antisymmetric matrices enters the EPR through a network-independent coupling coefficient $\Ker_{r,\ell}$ weighting the Frobenius overlap $\langle\mathcal A^{(r)},\mathcal A^{(\ell)}\rangle_F$ (Theorem~\ref{thm:kernel}). The EPR is thus controlled by the full hierarchy of walk nonreciprocity and its cross-length overlaps; a simple scalar non-normality measure such as the self-commutator norm $\|WW\T-W\T W\|_F$ cannot by itself determine the EPR.

Second, we recast this quadratic form in an equivalent representation that makes explicit how the network's structure shapes dissipation (Theorem~\ref{thm:counting}). Only two walk quantities enter the EPR expression: the aggregate weight of pairs of directed walks sharing both endpoints, and that of directed closed walks. This walk-counting form identifies network families in which the EPR simplifies.  For networks without closed walks, such as directed acyclic graphs (DAGs), only the endpoint-matched pairs survive. It also yields the EPR for nilpotent interactions, such as layered feedforward networks and oriented trees, where the spectral representation is unavailable.

Third, the two walk quantities introduced by the walk-counting form provide a direct bridge to standard spectral methods \cite{Godreche2019Characterising,Fyodorov2025Nonorthogonal,Nagayama2026Oscillatory}.  For diagonalizable $W$, we establish a spectral-walk correspondence (Theorem~\ref{thm:spectral}) showing that eigenvalues determine the directed closed-walk traces,  while biorthogonal eigenvector overlaps additionally determine the same-endpoint walk-pair traces; see~\eqref{eq:KL-spectral-main}. Non-orthogonal eigenvectors thus shape the walk-pair contribution to the EPR, but not the closed-walk contribution.	

%Nonzero nilpotent interactions, including nontrivial DAGs and feedforward networks, lie outside this diagonalizable spectral dictionary: all their eigenvalues vanish, although their powers and EPR need not.  Their EPR nevertheless remains exactly computable from the walk expansions of Theorems~\ref{thm:kernel} and~\ref{thm:counting}.

\begin{figure}[H]
  \centering
  \includegraphics[width=1\linewidth]{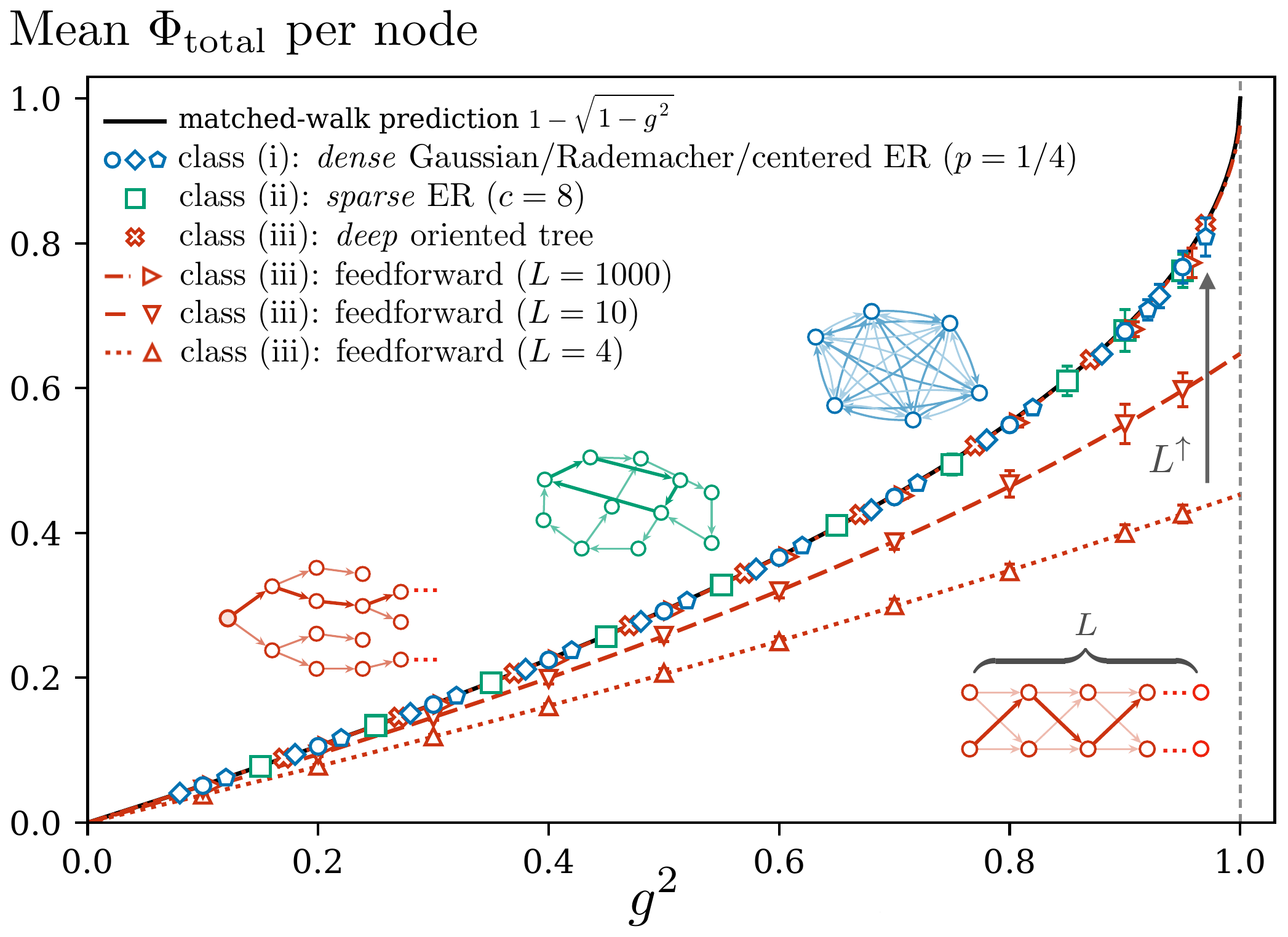}
\caption{{\bf The matched-walk prediction and the square-root law.} Numerical ensemble mean EPR per node (symbols) for acyclic networks and large-$N$ contractive recurrent ensembles agree with the analytic predictions shown in curves. The finite-depth feedforward curves follow Eq.~\eqref{eq:layered-main} and converge to the square-root law $\phi_*(g)$ as $L\to\infty$. The vertical dashed line at $g^2=1$ marks the radius-of-convergence boundary of the Catalan series resummation over all walk lengths, indicating the divergence of the susceptibility $\frac{d\phi_*}{d(g^2)}=\frac{1}{2\sqrt{1-g^2}}$ as $g\rightarrow 1^-$. Table~\ref{tab:ensemble-routes} summarizes the ensemble construction. Numerical details are in SM~M.}
  \label{fig:collapse}
\end{figure}

Finally, for large random interaction networks, the walk-counting formula reveals a universal square-root curve across recurrent ensembles and deep acyclic ensembles (Table~\ref{tab:ensemble-routes}); see Fig.~\ref{fig:collapse}.  By parametrizing each class by its asymptotic mean-squared interaction strength per node
\begin{equation}\label{eq:g2_para}
g^2:=\lim_{N\to\infty}\frac1N\EE_W\|W\|_F^2,
\end{equation}
we find that the mean EPR per node converges to
\begin{equation}\label{eq:sqr-limit}
\phi_*(g):=1-\sqrt{1-g^2}.
\end{equation}
These ensembles share three directed-walk properties that together produce this curve (Theorem~\ref{thm:matched-walk}).  First, \emph{matched-walk selection} makes two copies of the same directed walk the only leading walk-pair contribution at each fixed walk length; all other same-endpoint walk pairs and directed closed-walk contributions are subleading.  Second, the \emph{matched-walk moment condition}~\eqref{eq:matched-main} fixes the aggregate weight of the surviving length-$m$ pairs at $g^{2m}$ per node, so the diagonal coupling coefficient contributes $\Ker_{m,m}g^{2m}/2$ to the EPR at that length.  Finally, access to arbitrarily long walks, $m=1,2,\ldots$, resums these contributions into the square-root law~\eqref{eq:sqr-limit}.

The Supplementary Material (SM) verifies the matched-walk moment condition in the large-$N$ limit for each ensemble above. The resummation to $\phi_*(g)$ then follows once arbitrarily long walks are available. For finite feedforward networks with depth $L$, a depth-dependent correction produces the deviations from the square-root law as seen in Fig.~\ref{fig:collapse}, but vanishes as $L\to\infty$.

\paragraph{Setup and exact walk expansion.}---For $\bx\in\mathbb{R}^N$, we consider the OU dynamics
\begin{equation}
  \frac{d\bx}{dt}=-B\bx(t)+\bxi(t),\qquad B:=\Id-W,
\label{eq:MOU}
\end{equation}
where the entry $W_{ij}$ is the network interaction weighting the edge $j\to i$, and with isotropic noise $\langle\bxi(t)\bxi(t')\T\rangle=2\sigma\Id\,\delta(t-t')$. The noise isotropy assumption entails no loss of generality: an invertible whitening transformation makes any Gaussian white noise with a nondegenerate covariance isotropic without changing the EPR. Whitening does, however, transform the interaction matrix $W$, so the walk expansion and counting form must be constructed from the transformed matrix; see End Matter.

The OU dynamics converges to a unique stationary Gaussian distribution iff $-B$ is Hurwitz, equivalently $B=\Id-W$ is positive stable:
\begin{equation}
\min_{\lambda\in\operatorname{spec}(B)}\Re\lambda>0
\quad\Leftrightarrow\quad
\max_{\lambda\in\operatorname{spec}(W)}\Re\lambda<1.
\label{eq:stability-main}
\end{equation}
Positive stability makes the mean relax to zero and the covariance $S(t):=\operatorname{Cov}[\bx(t)]$ converge, independently of the initial state, to the unique positive-definite stationary covariance $S:=\lim_{t\to\infty}S(t)$. This stationary covariance satisfies the Lyapunov/Sylvester equation
\begin{equation}
BS+SB\T=2\sigma\Id.
\label{eq:Lyapunov-main}
\end{equation}
With the current-carrying antisymmetric Onsager matrix $Q:=BS-\sigma\Id=-Q\T$, the stationary EPR is~\cite{Gardiner2010Stochastic,Godreche2019Characterising,Tome2010}
\begin{equation}
\PhiTot=\frac1\sigma\Tr(QS^{-1}Q\T)
  =-\frac1\sigma\Tr(BQ)\ge0.
\label{eq:EPR-basics}
\end{equation}
Solving for $S$ is numerically straightforward, but substituting the numerical $S$ into~\eqref{eq:EPR-basics} obscures how the structure of $W$ generates dissipation~\cite{Gardiner2010Stochastic,Godreche2019Characterising}.

To reveal how $W$ enters the EPR, we expand the equivalent integral representation of the stationary covariance:
\begin{equation}
\begin{aligned}
S
&=2\sigma\int_0^\infty e^{-Bu}e^{-B\T u}\,du\\
&=\sigma\sum_{n,m\ge0}
\binom{n+m}{n}2^{-(n+m)}W^n(W\T)^m .
\end{aligned}
\label{eq:S-series-main}
\end{equation}
The first line holds throughout the positive-stability domain.  The series in the second line is exact under the stronger \emph{contractivity condition} $\rho(W)<1$, which guarantees absolute convergence. Throughout, we impose contractivity for the series expansion, not for stationarity: $B$ can be positive stable even when $\rho(W)\ge1$.

Under contractivity, substituting~\eqref{eq:S-series-main} into $Q=BS-\sigma\Id$ and~\eqref{eq:EPR-basics} yields an absolutely convergent power series for $\PhiTot$ in $W$.  The full series derivation for $S$, $Q$, and $\PhiTot$ are established in SM~C.  Note that since $S,Q\propto\sigma$, $\PhiTot$ becomes \emph{independent} of the scalar noise amplitude $\sigma$.

\paragraph{Main result 1.}---The EPR power series in $W$ simplifies immensely when expressed in terms of the walk-nonreciprocity matrices $\mathcal A^{(r)}$ introduced in~\eqref{eq:A-def}.  This exact reorganization yields the following quadratic form.

\begin{theorem}[Walk-nonreciprocity expansion]\label{thm:kernel}
For any real interaction matrix $W$ satisfying $\rho(W)<1$, the stationary EPR under isotropic noise is 
\begin{equation}
  \PhiTot=\sum_{r,\ell\ge1}\Ker_{r,\ell}
  \langle\mathcal A^{(r)},\mathcal A^{(\ell)}\rangle_F,
\label{eq:EPR-kernel}
\end{equation}
where $\langle \mathcal A^{(r)},\mathcal A^{(\ell)}\rangle_F:=\Tr(\mathcal{A}^{(r) \mathsf T} \mathcal A^{(\ell)})=\sum_{i,j}\mathcal A^{(r)}_{ij}\mathcal A^{(\ell)}_{ij}$ denotes the Frobenius inner product, and the graph-independent symmetric coupling coefficients are given by
\begin{equation}
\begin{aligned}
\Ker_{r,\ell}=\frac{1}{2^{r+\ell-1}}\!\bigg[&
2\binom{r+\ell-2}{r-1}-\binom{r+\ell-2}{r}\\[-2pt]
&-\binom{r+\ell-2}{r-2}\bigg].
\end{aligned}
\label{eq:kernel-def}
\end{equation}
On the diagonal, $\alpha_m:=\Ker_{m,m}=C_{m-1}/2^{2(m-1)}$, where $C_m=\frac{1}{m+1}\binom{2m}{m}$ is the $m$-th Catalan number.
\end{theorem}
The Catalan coefficient $\alpha_m$ is the fraction of the $2^{2(m-1)}$ possible one-dimensional random walks of length $2(m-1)$ starting at zero that remain nonnegative and end at zero (Dyck paths)~\cite{Stanley2015Catalan}; it is independent of the interaction graph. Off the diagonal, $\Ker_{r,\ell}$ couples unequal walk lengths, admits an interpretation as Lobb probabilities, and can have either sign; see SM~D\,3.  Positivity of the EPR is thus a property of the complete quadratic form, not of each nonreciprocity-pair contribution. 

For weak coupling, $\PhiTot(\varepsilon W)=\varepsilon^2\|W-W\T\|_F^2/4+O(\varepsilon^3)$, so the leading dissipation is determined by 
%the squared differences between couplings in opposite directions, $W_{ij}$ and $W_{ji}$, or equivalently by 
the squared norm of the nonreciprocal part $\mathcal A^{(1)}=(W-W\T)/2$ of $W$. This quadratic onset is characteristic of Onsager's linear near-equilibrium theory~\cite{Onsager1931}. Beyond weak coupling, each $\Ker_{r,\ell}$ weighs an alignment between walk-nonreciprocity patterns at two lengths and their full sum is the physical EPR.

\paragraph*{Main result 2.}---We now convert the walk-nonreciprocity representation of Theorem~\ref{thm:kernel} into a walk-counting form. Expanding each matrix power into weighted directed walks organizes the resulting products into two quantities: directed walk pairs sharing both endpoints and directed closed walks.  To define them precisely, let $\Gamma_r(j\!\to\! i)$ denote the directed $r$-step walks $\gamma=(i_0=j\to i_1\to\cdots\to i_r=i)$, allowing repeated vertices and edges, and define their weights by $w(\gamma):=\prod_{a=1}^r W_{i_a i_{a-1}}$.  Since $(W^r)_{ij}=\sum_{\gamma\in\Gamma_r(j\to i)}w(\gamma)$, these objects have, for $r,\ell,s\ge1$, the following trace and walk-sum representations:
\begin{equation}
\begin{aligned}
K_{r,\ell}
&:=\Tr[W^r(W\T)^\ell]
=\sum_{i,j}\ \sum_{\substack{\gamma\in\Gamma_r(j\to i)\\
                              \gamma'\in\Gamma_\ell(j\to i)}}
  w(\gamma)w(\gamma'),\\
L_s
&:=\Tr(W^s)
=\sum_i\ \sum_{\gamma\in\Gamma_s(i\to i)}w(\gamma).
\end{aligned}
\label{eq:KL-def}
\end{equation}
Combinatorially, $K_{r,\ell}$ is the total signed weight of ordered pairs of directed walks of lengths $r$ and $\ell$ that share both endpoints, whereas $L_s$ is the total signed weight of all closed $s$-step directed walks including those with repeated vertices or edges.
\begin{theorem}[Walk-counting form]\label{thm:counting}
For the nonreciprocal interaction network $W$ of Theorem~\ref{thm:kernel},
\begin{equation}
  \PhiTot=\frac12\sum_{r,\ell\ge1}\Ker_{r,\ell}K_{r,\ell}
  -\frac12\sum_{s\ge2}2^{-(s-2)}L_s.
\label{eq:EPR-counting}
\end{equation}
\end{theorem}
This follows from substituting $\langle\mathcal A^{(r)},\mathcal A^{(\ell)}\rangle_F =\tfrac12(K_{r,\ell}-L_{r+\ell})$ into the quadratic form of Theorem~\ref{thm:kernel}, grouping the closed-walk terms by their total length $s=r+\ell$, and using $\sum_{r=1}^{s-1}\Ker_{r,s-r}=2^{-(s-2)}$.
 Conveniently, the counting form makes several architectural simplifications immediate. Any directed acyclic graph (DAG) has no directed closed walks, so $L_s=0$ for all $s\ge1$, leaving only same-endpoint walk pairs. A stronger simplification occurs for directed chains, layered feedforward networks, and rooted oriented trees; their walks sharing both endpoints must have the same length, so $K_{r,\ell}=0$ for $r\ne\ell$. Thus, 
\begin{equation}\label{eq:DAG-same-endpoint-EPR}
\PhiTot=\frac12\sum_{m\ge1}\alpha_m K_{m,m}
=\frac12\sum_{m\ge1}\alpha_m\|W^m\|_F^2.
\end{equation}

For a uniform directed chain (DC) with the interaction strength $g$, $K_{m,m}=(N-m)g^{2m}$ and $\|W\|_F^2/N\to g^2$. Thus, for $0\le g<1$, we obtain, deterministically, the square-root law:
\begin{equation}\label{eq:dc-main}
\lim_{N\to\infty}\frac{\PhiTot^{\text{DC}}}{N}
=\frac12\sum_{m\ge1}\alpha_m g^{2m}
=1-\sqrt{1-g^2}
=\phi_*(g).
\end{equation}
This recovers the result of the large, totally asymmetric open chain in Ref.~\cite{Godreche2019Characterising}; see SM~F and~K.

\paragraph*{Main result 3.}---Existing non-Hermitian spectral formulas express the EPR from the eigenvalues and biorthogonal eigenvector overlaps of $B = \Id - W$~\cite{Godreche2019Characterising,Fyodorov2025Nonorthogonal}, but leave their connection to the network's walk structure implicit. Since $B$ and $W$ share eigenvectors and their eigenvalues differ only by a shift, we formulate the exact correspondence in terms of $W$.  For diagonalizable $W$, let $Wv_a=\lambda_av_a$, $u_a\T W=\lambda_au_a\T$, and $u_a\T v_b=\delta_{ab}$, and define the bilinear overlap
\begin{equation}
  O_{ab}:=(u_a\T u_b)(v_b\T v_a).
\label{eq:O-def}
\end{equation}
\begin{theorem}[Spectral-walk correspondence]\label{thm:spectral}
For every real $W$ diagonalizable over $\mathbb C$ and all $r,\ell,s\ge1$,
\begin{align}
  K_{r,\ell}&=\sum_{a,b}\lambda_a^r\lambda_b^\ell O_{ab},\nonumber\\[-2pt]
  L_s&=\sum_a\lambda_a^s,\label{eq:KL-spectral-main}\\[-2pt]
  \langle\mathcal A^{(r)},\mathcal A^{(\ell)}\rangle_F
  &=-\frac14\sum_{a,b}(\lambda_a^r-\lambda_b^r)
  (\lambda_a^\ell-\lambda_b^\ell)O_{ab}.\nonumber
\end{align}
\end{theorem}
This correspondence reveals that eigenvalues alone determine the closed-walk trace $L_s$, whereas the same-endpoint walk-pair trace $K_{r,\ell}$ also probes eigenvector geometry. Substituting~\eqref{eq:KL-spectral-main} into~\eqref{eq:EPR-counting} recovers, on $\rho(W)<1$, the known eigenvalue-overlap EPR formula\footnote{Although its recovery from the series expansion requires $\rho(W)<1$, Eq.~\eqref{eq:Phi-spectral} itself holds throughout the diagonalizable OU-stable domain $\max_{\lambda\in\operatorname{spec}(W)}\Re\lambda<1$.}
\begin{equation}
  \PhiTot=-\frac12\sum_{a,b}
  \frac{(\lambda_a-\lambda_b)^2}{2-\lambda_a-\lambda_b}O_{ab}.
\label{eq:Phi-spectral}
\end{equation}
It follows from the biorthogonal \(Q\)-representation and EPR trace identity~\cite{Godreche2019Characterising} after symmetrization, and is equivalent, after conjugate-index relabeling, to the overlap formula of Fyodorov \textit{et al.}~\cite{Fyodorov2025Nonorthogonal}.  The new result here is formula~\eqref{eq:KL-spectral-main}, which connects these spectral quantities to directed-walk statistics.

%\paragraph*{Main result 4.}---\tc{The square-root curve $\phi_*(g)$ appears for the uniform directed chain~\cite{Godreche2019Characterising} and, spectrally, for large Ginibre matrices~\cite{Fyodorov2025Nonorthogonal}; see End Matter.  The acyclic ensembles sharpen this contrast: for every realization and every $g$, $W$ is nilpotent, with $\operatorname{spec}(W)=\{0\}$ and $\rho(W)=0$.  The eigenvalues therefore cannot encode the EPR's $g$ dependence. For nonzero $W$, the complete biorthogonal eigenbasis required by Main result~3 is absent, whereas the walk-counting formula of Theorem~\ref{thm:counting} remains exact and terminates.  The condition below isolates the matched-walk hierarchy shared with the recurrent ensembles.}

\paragraph*{Main result 4.}---The square-root curve $\phi_*(g)$ in~\eqref{eq:sqr-limit} appears explicitly for the deterministic uniform directed chain~\eqref{eq:dc-main}~\cite{Godreche2019Characterising} and, after reparametrization, in a seemingly unrelated non-Hermitian spectral calculation for a random Ginibre ensemble~\cite{Fyodorov2025Nonorthogonal}; see End Matter. We now identify their common \emph{matched-walk} structure. Together with the walk-counting form, it determines the common Catalan generating function, giving a single derivation of the square-root law across disparate ensembles---including nilpotent acyclic networks, where the spectral representation is unavailable.

%\tc{Consider the pairs of directed walks counted by $K_{m,\ell}$: the two walks share their starting and ending nodes but may have different lengths or follow different routes.  We say that an ensemble exhibits \emph{matched-walk selection} when, at every fixed length $m$, only pairs that traverse exactly the same directed edges in the same order contribute at leading order in the large-$N$ ensemble average.  Unequal-length pairs and directed closed-walk traces are likewise subleading.  This motivates the following ensemble-level sufficient condition.}

\begin{theorem}[Square-root law resummation under the matched-walk
moment condition]
\label{thm:matched-walk}
Let $\{W_N\}_{N\ge1}$ be a sequence of random $N\times N$ interaction matrices with asymptotic mean-squared interaction strength per node $g^2$ as defined in Eq.~\eqref{eq:g2_para}.  Suppose that the sequence satisfies the \emph{matched-walk moment condition} with parameter
$g$, namely, for every fixed $m,\ell\ge1$ and $s\ge2$,
\begin{equation}
\lim_{N\to\infty}\frac1N\EE K_{m,\ell}=\delta_{m\ell}g^{2m},
\qquad
\lim_{N\to\infty}\frac1N\EE L_s=0.
\label{eq:matched-main}
\end{equation}
For a fixed walk-length cutoff $M$, define
\begin{align}
\Phi_M
&:=\sum_{1\le r,\ell\le M}\Ker_{r,\ell}
\langle\mathcal A^{(r)},\mathcal A^{(\ell)}\rangle_F \nonumber \\
&=\frac12\sum_{1\le r,\ell\le M}
\Ker_{r,\ell}\bigl(K_{r,\ell}-L_{r+\ell}\bigr).
\label{eq:PhiM-main}
\end{align}
Then, for every fixed $M$,
\begin{equation}
\lim_{N\to\infty}\frac{\EE\Phi_M}{N}
=\frac12\sum_{m=1}^{M}\alpha_mg^{2m}
\mathrel{\mathop{\to}\limits_{\mathclap{\scriptscriptstyle M\to\infty}}}
\phi_*(g)=1-\sqrt{1-g^2}.
\label{eq:universal-intro}
\end{equation}
\end{theorem}

The condition~\eqref{eq:matched-main} is sufficient. In recurrent ensembles, directed walk counting reveals that, at leading order, only pairs comprising two copies of the same directed walk survive, whereas the nilpotent acyclic ensembles realize its deep-network limit through exact terminating counts. Once this moment condition is established, the resulting Catalan-weighted series resums to $\phi_*(g)$. 

The moment condition is not automatic. In the elliptic ensemble of Sommers \emph{et al.}~\cite{SommersCrisanti1988}, normalize the centered off-diagonal entries by $\EE W_{ij}^2=g^2/N$ and write their reciprocal correlation as $\EE(W_{ij}W_{ji})=\tau g^2/N$ for $i\ne j$.  Thus $\lim_{N\to\infty}\tfrac1N\EE\|W\|_F^2= g^2$, whereas the two-step closed-walk trace $L_2=\sum_{i,j}W_{ij}W_{ji}$ gives $\lim_{N\to\infty}\tfrac1N\EE L_2=\tau g^2 \neq 0$ for $\tau\ne0$.  Hence, this elliptic ensemble lies outside~\eqref{eq:matched-main}.

\raggedbottom
\paragraph*{Acyclic ensembles: square-root law with all eigenvalues fixed at zero.} For a layered feedforward ensemble with depth $L$, SM~K\,1 gives the exact finite-depth mean
\begin{equation}
  \lim_{N\to\infty}\frac{\EE\PhiTot}{N}=\frac12\sum_{m=1}^{L-1}
  \alpha_m\Bigl(1-\frac mL\Bigr)g^{2m}.
\label{eq:layered-main}
\end{equation}
For every finite $L$, the right-hand side is an analytic polynomial valid for every finite $g$, while nilpotency gives $\rho(W)=0$ independently of
$g$.  For $0\le g\le1$ it converges to $\phi_*(g)$ as $L\to\infty$, whereas for $g>1$ it diverges with depth despite every finite network remaining OU stable.  Rooted out-trees have an analogous finite-depth formula and converge to $\phi_*(g)$ for $0\le g\le1$ when the depth grows large; see SM~K\,2.

\paragraph*{Recurrent ensembles.} Asymptotic matched-walk counting proves the matched-walk moment condition for dense iid and sparse directed Erd\H{o}s--R\'enyi matrices under prescribed ensemble statistics in Table~\ref{tab:ensemble-routes}; see SM~I and~J.  For centered dense ensembles at fixed $g<1$, spectral-radius convergence implies that $\rho(W_N)<1$ with probability tending to one, so restricting to contractive samples leaves these fixed-length coefficients unchanged~\cite{BordenaveChafaiGarciaZelada2022,AltErdosKruger2021}; the same convergence places the Catalan convergence boundary $g=1$ at \emph{both} the typical large-$N$ contractivity and OU stability boundary of these dense ensembles.

\begin{figure}[H]
  \centering
  \includegraphics[width=0.95\linewidth]{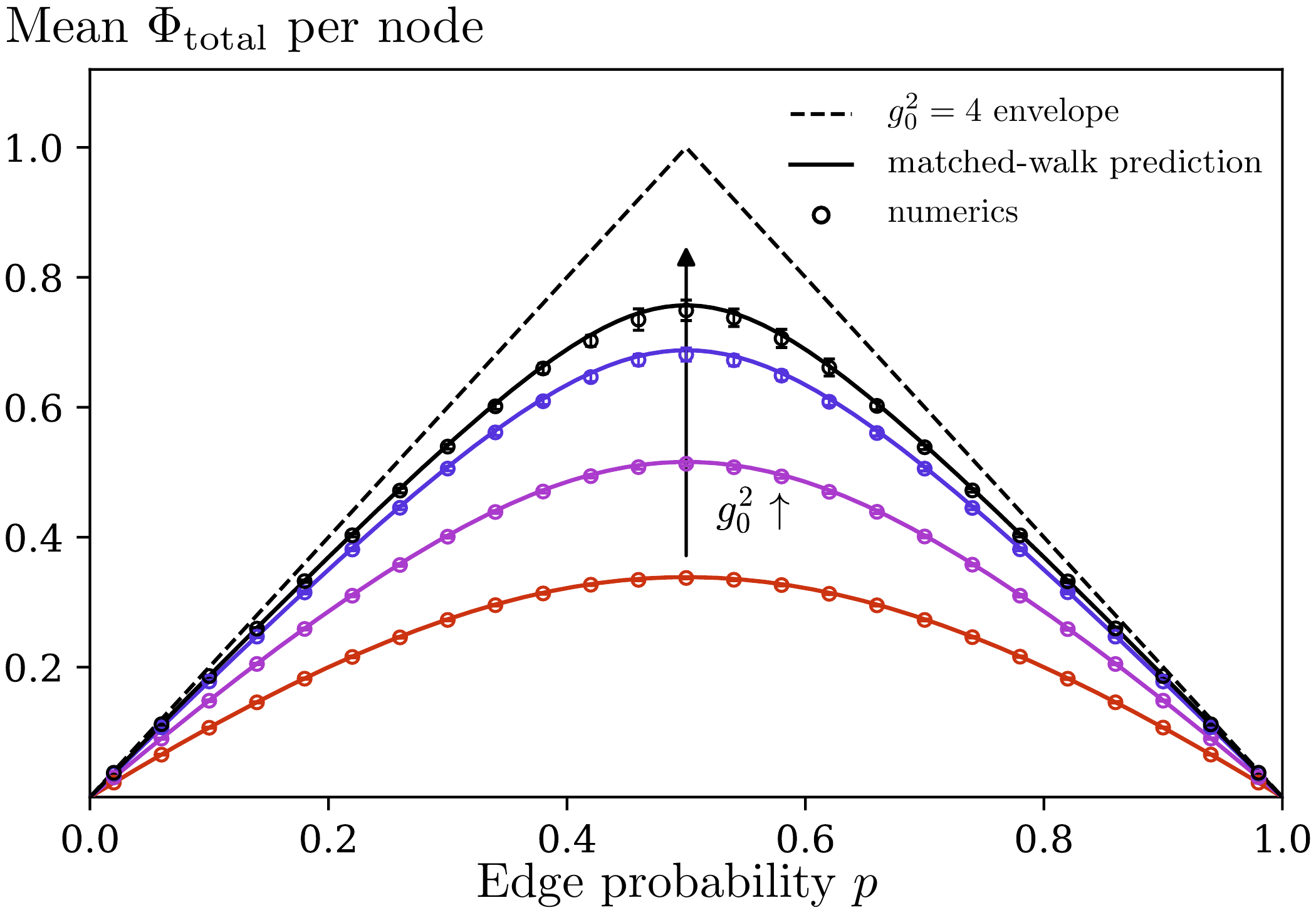}
\caption{{\bf Dissipation is non-monotonic in edge probability.} For centered dense Erd\H{o}s-R\'enyi networks the effective interaction is $g^2=g_0^2p(1-p)$, so the EPR rises then falls with edge probability, peaking at $p=1/2$: beyond half-filling, adding edges dilutes the effective interaction strength and lowers dissipation. The square-root law $\phi_*(g)$ predicts the full EPR (solid lines) at fixed $g_0^2\in\{1.50^2,1.75^2,1.90^2,1.94^2\}$. Symbols are large-$N$ ensemble means from direct Lyapunov solution, conditioned on $\rho(W)<1$, with error bars showing SD across retained realizations. The dashed $g_0^2=4$ envelope touches the Catalan-series convergence boundary $g^2=1$ at $p=1/2$; see SM~M. This is an analytic benchmark for the edge-probability dependence reported numerically in~\cite{Kaluarachchi2024Broken}.
}
  \label{fig:er}
\end{figure}

\noindent We do not prove the corresponding conditioning result for sparse networks, and instead test the complete EPR numerically over contractive samples. The resulting square-root prediction agrees with the independent large-$N$ Ginibre calculation~\cite{Fyodorov2025Nonorthogonal} (End Matter) and with the numerical ensemble means in Figs.~\ref{fig:collapse} and~\ref{fig:er} (SM~H).

\paragraph{Discussion and outlook.}---Directed walk properties determine the steady-state EPR in nonreciprocally coupled systems even when the biorthogonal spectral representation is unavailable, as in nilpotent acyclic ensembles whose eigenvalues remain fixed at zero. Theorem~\ref{thm:matched-walk} gives a sufficient matched-walk moment condition for ensemble resummation to $\phi_*(g)$. 
Identifying further ensembles that satisfy this condition and having access to arbitrarily long walks would extend the class collapsing onto the square-root law beyond Table~\ref{tab:ensemble-routes}. 

The moment condition\mn{~\eqref{eq:matched-main}} suggests where multiparameter behavior can enter; through surviving
$N^{-1}\EE K_{m,\ell}$ with $m\ne\ell$  or $N^{-1}\EE L_s$ for $s\ge2$.  The elliptic ensemble is an example~\cite{SommersCrisanti1988}: $\EE(W_{ij}W_{ji})=\tau g^2/N$ gives $N^{-1}\EE\PhiTot=(1-\tau)g^2/2+O(g^4)$, departing from $\phi_*(g)$ at leading order for  $\tau\ne0$.  Sparse ensembles with finite mean degree, heterogeneous timescales, interaction-sign constraints, or block and trophic structure are worth further investigations~\cite{Krumbeck2021Fluctuation,Mambuca2022Dynamical,Chotibut2025}. 

Real systems are nonlinear, but near a stable steady state their fluctuations obey OU dynamics with $W = \Id + J$, where $J$ is the Jacobian  around a stable operating point (End Matter), so our results apply to local EPR. These situations include, for example, sparse structured neural networks that sustain a NESS for working memory computation~\cite{Rungratsameetaweemana2025,Chotibut2025}, and random neural networks near the onset of chaos~\cite{Pham2025Irreversibility}, whose weak-coupling limit agrees with $\phi_*(g)$ (End Matter). 

Ultimately, the square-root law shows that topologically and spectrally distinct networks can share the same EPR through common  directed-walk statistics, revealed through the walk-counting formulation. These results potentially offer an analytical route for connecting interaction network structure to thermodynamic constraints on information processing, learning, and computation in biological and artificial networks~\cite{Parrondo2015Thermodynamics,Goldt2017Stochastic,Wolpert2019Stochastic,Aifer2024Thermodynamic,Melanson2025Thermodynamic}.

\begingroup
\renewcommand{\addcontentsline}[3]{}
\paragraph*{Acknowledgement.}---
We thank Wojciech Tarnowski, Oleg Evnin, Yizhuang Liu and Weerawit Horinouchi for useful discussions, and Eytan Katzav for pointing out the connection to Lobb numbers. TC acknowledges funding support from the NSRF via the Program Management Unit for Human Resources \& Institutional Development, Research and Innovation [Grant No.\ B39G690073]. EGN and MAN are supported by the Priority Research Area DigiWorld under the program Excellence Initiative - Research University at the Jagiellonian University in Kraków. The authors conceived the research, performed the analytical and numerical work, and drafted the manuscript. Beginning in mid-2026, Claude and ChatGPT were used for mathematical verification, language editing, and code cleanup. The authors verified all AI-assisted output and take full responsibility for the manuscript.

\paragraph*{Data and code availability.}---A self-contained Python implementation, together with the data needed to reproduce the figures, will be available in a public repository at \url{https://github.com/TChotibut/epr-walk-counting-code}. A 50-page Supplemental Material document containing detailed derivations, proofs, and numerical protocols has been submitted with the manuscript for peer review. It is not included in this public preprint while the manuscript is under review and will be made publicly available following peer review.
\endgroup

% =============================================================
% END MATTER
% =============================================================
\begingroup
\renewcommand{\addcontentsline}[3]{}
\section*{End Matter}
\endgroup

\phantomsection
\paragraph{Appendix A: Generality of the OU dynamics.}
\label{app:linearizations}---
Consider a general nonlinear system with additive Gaussian noise, $\dot{\bx}=\boldsymbol F(\bx)+\bxi(t)$, near a linearly stable deterministic fixed point $\bx^*$.  We use the dimensionless-time convention and define the resulting dimensionless Jacobian by $J_{ij}:=\left.\partial F_i/\partial x_j\right|_{\bx^*}$.  The small fluctuation $\delta\bx=\bx-\bx^*$ then obeys an OU dynamics of Eq.~\eqref{eq:MOU} with $B=-J=\Id-W.$ The interaction matrix is read directly from $J$ as
\[W=\Id+J.
\label{eq:linearized-effective-W}
\]
Many dynamical systems admit the additive pairwise interaction form $F_i(\bx)=R_i(x_i)+\sum_{j\ne i}A_{ij}h(x_i,x_j)$, where $R_i$ describes
the intrinsic dynamics of node $i$ and $h$ describes the contribution of
node $j$ to node $i$.  The Jacobian entries read
$J_{ij}=A_{ij}\partial_2h(x_i^*,x_j^*)$ and hence
\begin{equation}
W_{ij}=A_{ij}\partial_2h(x_i^*,x_j^*),\qquad i\ne j.
\end{equation}
Table~\ref{tab:linearizations} gives representative examples.  The stationary Gaussian fluctuation state is a NESS whenever the linearized
dynamics breaks detailed balance.  Our results describe the local EPR of Gaussian fluctuations near the fixed point, not the full nonlinear EPR
away from the fixed point.  Linear stability makes $B=-J$ positive stable; the convergent
walk expansion additionally requires $\rho(W)=\rho(\Id+J)<1$.

\begin{widetext}
\begin{center}
\refstepcounter{table}\label{tab:linearizations}
\begin{minipage}{0.92\textwidth}
\small
\textbf{TABLE~\thetable.} Representative models and their translations from additive pairwise nonlinear dynamics to the off-diagonal interaction
$W_{ij}=J_{ij}$ near a stable fixed point $\bx^*$.  
\end{minipage}
\par\vspace{0.5em}
\begingroup
\setlength{\tabcolsep}{5pt}
\renewcommand{\arraystretch}{1.2}
\begin{ruledtabular}
\begin{tabular}{lcc}
\bf{Model} & \bf{Additive pairwise interaction} $h(x_i,x_j)$ & $W_{ij}\ (i\ne j)$\\
\hline
Rate-based neural network~\cite{Sompolinsky1988,Dayan2005,Kadmon2015}
& $\tanh x_j$
& $A_{ij}\operatorname{sech}^2(x_j^*)$ \\
DeGroot consensus formation~\cite{Friedkin1990,Olfati-saber2007,Castellano2009,Vaidya2021}
& $x_j-x_i$
& $A_{ij}$ \\
Lotka-Volterra~\cite{Bunin2017,Galla2018Eco}
& $x_i x_j$
& $A_{ij}x_i^*$ \\
Kuramoto-Sakaguchi synchronization~\cite{Sakaguchi1988,Childs2008}
& $\sin(x_j-x_i-\alpha)$
& $A_{ij}\cos(x_j^*-x_i^*-\alpha)$\\
\end{tabular}
\end{ruledtabular}
\endgroup
\end{center}
\end{widetext}

\phantomsection
\paragraph{Appendix B: Generality of the isotropic-noise convention.}
\label{app:diffusion-scope}---
The main text uses the constant diffusion matrix $D=\sigma\Id$, so $\langle\bxi(t)\bxi(t')\T\rangle=2\sigma\Id\,\delta(t-t')$.  More generally, any constant diffusion matrix $D\succ0$ can be whitened.  With its unique symmetric positive-definite square root~\cite{Higham2008Functions}, set
\begin{equation}
\begin{aligned}
\by&=D^{-1/2}\bx,\\
B_D&=D^{-1/2}BD^{1/2}=\Id-W_D,\\
W_D&=D^{-1/2}WD^{1/2}.
\end{aligned}
\label{eq:whitening-endmatter}
\end{equation}
Then $\by$ obeys the OU dynamics $\dot\by=-B_D\by+\bxi_D(t)$, where $\bxi_D(t):=D^{-1/2}\bxi(t)$ has covariance $\langle\bxi_D(t)\bxi_D(t')\T\rangle=2\Id\,\delta(t-t')$. The stationary EPR is unchanged~\cite{Godreche2019Characterising}.  Subject to the stated hypotheses, the exact formulas retain their form with $W\mapsto W_D$; in particular, the exact EPR walk expansion is justified here under $\rho(W_D)=\rho(W)<1$. Because $W_D$ is similar to $W$, whitening preserves eigenvalues, spectral radius, and closed-walk traces, but not generally the Frobenius norm or the walk-pair traces $K_{r,\ell}$.  To apply the square-root prediction, the \emph{whitened} ensemble must satisfy the matched-walk moment condition with coupling scale
$
g_D^2:=\lim_{N\to\infty}\frac1N\EE\|W_D\|_F^2.
$
The coordinate, covariance, EPR-invariance, and detailed-balance transformations are given in SM~B, while the transformed walk and random-ensemble quantities are collected in SM~L.

\phantomsection
\paragraph{Appendix C: Ensemble parametrization and the common coupling strength.}
\label{app:scope}---
The ensembles compared in Fig.~\ref{fig:collapse} are listed in
Table~\ref{tab:ensemble-routes}. They share the mean-squared coupling
strength per node in the large-$N$ limit. Namely, 
\begin{equation}
g_N^2:=\frac1N\EE\|W\|_F^2 \to g^2.
\label{eq:ensemble-coupling-endmatter}
\end{equation}
For the four recurrent ensembles and the out-tree,
$g_N^2=(1-N^{-1})g^2\to g^2$ as $N\to\infty$. For an $L$-layer
feedforward network, $g_N^2=(1-L^{-1})g^2\to g^2$ in the deep-network
limit $L\to\infty$. The last column of
Table~\ref{tab:ensemble-routes} lists the corresponding SM section containing the matched-walk analysis and limiting square-root resummation result.

\begin{table*}[tbp]
\caption{Ensembles used in Fig.~\ref{fig:collapse}. Each realizes the common asymptotic mean-squared interaction strength per node $g^2$ in Eq.~\eqref{eq:ensemble-coupling-endmatter}. The last column lists the corresponding section number in the SM.}
\label{tab:ensemble-routes}
\centering
\footnotesize
\setlength{\tabcolsep}{3pt}
\renewcommand{\arraystretch}{1.17}
\begin{ruledtabular}
\begin{tabular}{p{0.15\textwidth}p{0.66\textwidth}p{0.11\textwidth}}
\bf{Ensemble} & \bf{Construction} & \bf{Details}\\
\hline
Real Gaussian & $W_{ij}=gZ_{ij}/\sqrt N$ for $i\ne j$, $Z_{ij}\stackrel{\rm iid}{\sim}\mathcal N(0,1)$, $W_{ii}=0$ & SM~I\\
Rademacher & $W_{ij}=g\varepsilon_{ij}/\sqrt N$, $\mathbb P(\varepsilon_{ij}=\pm1)=1/2$, $W_{ii}=0$ & SM~I\,4\\
Centered dense ER & $W_{ij}=g(A_{ij}-p_*)/\sqrt{Np_*(1-p_*)}$ for $i\ne j$, $A_{ij}\sim{\rm Bernoulli}(p_*)$, $W_{ii}=0$; $p_*=1/4$ in Fig.~\ref{fig:collapse} & SM~I\,3\\
Sparse directed ER & $W_{ij}=A_{ij}\xi_{ij}$ for $i\ne j$, with $A_{ij}\stackrel{\rm iid}{\sim}{\rm Bernoulli}(c/N)$; the $\xi_{ij}$ are iid, independent of the $A_{ij}$, and satisfy $\EE\xi_{ij}=0$ and $\EE\xi_{ij}^2=g^2/c$.  Fig.~\ref{fig:collapse} uses the mean degree $c=8$ and $\xi_{ij}\sim\mathcal N(0,g^2/c)$; $W_{ii}=0$ & SM~J\\
Layered feedforward & $N=Ln$: $L$ ordered layers $V_1,\ldots,V_L$ of width $n$; the only nonzero blocks are the forward blocks $V_\ell\to V_{\ell+1}$,
$\widetilde G_\ell:=\widetilde W_{\ell+1,\ell}=gX_\ell/\sqrt n$ for $\ell=1,\ldots,L-1$, where $X_\ell\in\RR^{n\times n}$ and each element in $\{(X_\ell)_{ij}\}$ is iid $\mathcal N(0,1)$ & SM~K\,1\\
Oriented out-tree &
Rooted tree with edges directed parent-to-child and $W_e=g\xi_e$; the $\xi_e$ are iid, and $\EE\xi_e^2=1$.  Fig.~\ref{fig:collapse} uses uniform Cayley trees with $|\xi_e|=1$ &
SM~K\,2\\
\end{tabular}
\end{ruledtabular}
\end{table*}

\phantomsection
\paragraph{Appendix D: Connection to random neural networks.}\label{app:pham}---
Pham, Alonso, and Proesmans study a random recurrent rate network in which each unit applies a nonlinear activation function to its total synaptic input~\cite{Pham2025Irreversibility}:
\begin{equation}
\begin{aligned}
\dot x_i&=-x_i+\tanh\!\left(\sum_{j\ne i}J_{ij}x_j\right)+\xi_i(t),\\
J_{ij}&\sim\mathcal N(0,J^2/N).
\end{aligned}
\label{eq:pham-endmatter}
\end{equation}
Here $J_{ij}$ denotes the random coupling matrix, not the Jacobian $J$ in Appendix A.  Writing the total synaptic input to unit $i$ as $u_i=\sum_{j\ne i}J_{ij}x_j$, this model applies $\tanh$ after the inputs have been summed.  This differs from the rate model in the first row of Table~\ref{tab:linearizations}, where $\tanh(x_j)$ transforms each presynaptic state before its weighted contribution is summed.  At a general fixed point, the off-diagonal linearized interaction is $W_{ij}=J_{ij}\operatorname{sech}^2(u_i^*)$.  At the silent fixed point, $W_{ij}=J_{ij}$, and since $\EE J_{ij}^2=J^2/N$, the zero-diagonal Gaussian ensemble in Table~\ref{tab:ensemble-routes} applies with $g=J$.  The corresponding large-$N$ mean EPR per neuron of the linearized OU dynamics is
\begin{equation}
\phi_*(J)=1-\sqrt{1-J^2}
=\frac{J^2}{2}+O(J^4),
\qquad 0\le J<1.
\label{eq:pham-square-root-endmatter}
\end{equation}
Its leading term, $J^2/2$, agrees with the weak-coupling expansion of the nonlinear theory in Ref.~\cite{Pham2025Irreversibility}.  Moreover, $\phi_*(J)\to1$ as $J\to1^-$.  This limiting value agrees with the nonlinear theory's EPR at the onset of chaos in the low-noise limit, where its critical coupling approaches $J_c\to1$.  At finite noise, however, $J_c>1$, and the complete nonlinear EPR is not identified with the OU square-root curve. 

Separately, Fyodorov \emph{et al.} study the large-$N$ linear neural network model
$\dot{\bx}=(-\mu\Id+X+\nu M)\bx+\bxi(t)$, where $X$ is a unit-radius Ginibre matrix with a row-sum constraint and $\nu M$ is a rank-one Rajan-Abbott structure. Non-Hermitian spectral formulas and eigenvector overlap calculations give  $\lim_{N\to\infty}N^{-1}\EE\PhiTot=(1+\nu^2)/(\mu+\sqrt{\mu^2-1})$ for $\mu>1$~\cite{Fyodorov2025Nonorthogonal}.   At $\nu=0$, the dimensionless time $t'=\mu t$ gives $W=X/\mu$ and hence $g^2=\lim_{N\to\infty}N^{-1}\EE\|W\|_F^2=\mu^{-2}$; the EPR is a rate, so it rescales as $\PhiTot\mapsto\mu^{-1}\PhiTot$, and the rescaled mean EPR per node becomes
$1-\sqrt{1-g^2}$.  Their stability condition $\mu>1$ is precisely the Catalan convergence domain $g<1$.  Thus the spectral calculation contains
the same square-root law in a different time normalization, while our walk-counting form identifies its origin in matched-walk selection and Catalan resummation.

% =============================================================
% Main bibliography
% =============================================================
\bibliographystyle{apsrev4-2}
\let\savedaddcontentsline\addcontentsline
\renewcommand{\addcontentsline}[3]{}
\bibliography{epr_prl}
\let\addcontentsline\savedaddcontentsline

\end{document}